\documentclass[prb,twocolumn,showpacs,amsmath,amssymo]{revtex4}
\usepackage{graphicx}
\usepackage{dcolumn}
\usepackage{bm}

\begin{document}
\title{Thermodynamic properties of the itinerant-boson ferromagnet}

\author{Chengjun Tao, Peilin Wang, Jihong Qin, and Qiang Gu}

\affiliation{Department of Physics, University of Science and Technology Beijing,
Beijing 100083, P.R. China}

\date{\today}

\begin{abstract}
Thermodynamics of a spin-1 Bose gas with ferromagnetic interactions
are investigated via the mean-field theory. It is apparently shown
in the specific heat curve that the system undergoes two phase
transitions, the ferromagnetic transition and the Bose-Einstein
condensation, with the Curie point above the condensation
temperature. Above the Curie point, the susceptibility fits the
Curie-Weiss law perfectly. At a fixed temperature, the reciprocal
susceptibility is also in a good linear relationship with the
ferromagnetic interaction.
\end{abstract}

\pacs{75.40.Cx, 75.10.Lp, 75.30.Kz, 03.75.Mn}

\maketitle

\section{Introduction}
The realization of spinor Bose-Einstein condensation in optical
traps\cite{ketterle,barrett} has stimulated enormous interest in
magnetic properties of quantum Bose
gases\cite{burke,ho,ohmi,isoshima,ueda,gu1,szirmai,zhang,gu2,
mur,schma,chang,Berkeley}. In optical traps, the hyperfine degree of
freedom of confined atoms, such as $\rm ^{87}Rb$, is released and
therefore the atom can exhibit magnetism. More intriguingly, an
exchange-like spin-spin interaction can be present between atoms. In
the $F=1$ $\rm ^{87}Rb$ atoms, the interaction is
ferromagnetic\cite{burke}, so the $\rm ^{87}Rb$ gas appears to be a
prototype of itinerant-boson
ferromagnet\cite{ho,ohmi,isoshima,ueda,gu1}.

Ferromagnetism is one of the central research themes in condensed
matter physics~\cite{mohn,moriya}. Two types of ferromagnetism have
already been intensively studied: local-moment ferromagnetism and
itinerant-electron ferromagnetism. Although particles in these two
systems obey different statistics, they both share some common
features. For example, both ferromagnets have a Curie point, above
which the susceptibility conforms to Curie-Weiss law. Nonetheless,
from the theoretical point of view, the origin of Curie-Weiss law is
quite different for these two systems. In insulators it is due to
local thermal spin fluctuations and can be easily explained in the
mean-field approximation. On the other hand, in itinerant-electron
ferromagnets the Curie-Weiss law may be caused by the mode-mode
coupling between spin fluctuations and the theoretical treatment is
much more complicated~\cite{moriya}. An appropriate theory is the
self-consistent renormalization (SCR) theory~\cite{murata} which
goes beyond the Hartree-Fock approximation and the random-phase
approximation. The SCR theory succeeds in explaining various
magnetic properties of itinerant-electron ferromagnets and is also
extended to treat the specific heat~\cite{taka}.

The $\rm ^{87}Rb$ gas provides opportunity to study the third type
of ferromagnetism. Ho~\cite{ho}, Ohmi and Machida~\cite{ohmi} have
studied its ground state properties and the spin-wave spectrum. The
long wavelength spectrum is linear in ${\bf k}$, the wave vector, as
in the two former cases. In our previous papers, we have
investigated the finite-temperature properties, especially the Curie
point~\cite{gu1}. We suggest that the phase diagram in itinerant
bosons should be more complicated than the other two ferromagnets,
because the Bose system has an intrinsic phase transition, other
than the ferromagnetic transition. An interesting conclusion we
arrived is that its Curie point, $T_F$, is never below the
Bose-Einstein condensation temperature, $T_C$, regardless of the
magnitude of the ferromagnetic coupling~\cite{gu1}. Kis-Szabo {\it
et al} got the same point later~\cite{szirmai}. However,
thermodynamics of the itinerant-boson ferromagnet has not yet been
investigated systematically so far.

The purpose of this paper is to calculate the thermodynamic
quantities of ferromagnetic bosons. As in the fermion case, the
specific heat and magnetic susceptibility are of the most interest.
In Section 2, we introduce the mean-field approximation to deal with
ferromagnetic interaction, taking the spin-1 Bose gas as an example.
In Section 3, phase transitions are discussed by calculating the
free energy and specific heat. In Section 4, the susceptibility
above the Curie point is calculated. A summary is given in the last
section.

\section{The Mean-field Approximation}
The spin-1 Bose gas with ferromagnetic couplings is described by the
following Hamiltonian,
\begin{eqnarray}\label{e01}
\hat H &=& \sum_\sigma \int d{\bf r} \hat \psi^{\dag}_\sigma({\bf r})
    \left( \frac 1{2m}\nabla^2 - \sigma h_e \right) \hat \psi_\sigma({\bf r}) \nonumber\\
    &&- \frac 12 I_s  \int d{\bf r} \hat {\bf S}({\bf r})\cdot \hat {\bf S}({\bf r})   ,
\end{eqnarray}
where $\hat \psi_\sigma({\bf r})$ is the quantum field operator for
annihilating an atom in spin state $|\sigma\rangle$ at site $r$. For
a spin-1 gas, $\sigma= +1, 0, -1$. The parameter $h_e$ denotes the
external magnetic field. The last term represents the ferromagnetic
exchange between two different bosons meeting at site $r$ and
$I_s(>0)$ is the exchange constant. $\hat{\bf S} =\{\hat S^x, \hat
S^y, \hat S^z\}$ are the spin operators, which can be expressed via
the $3\times3$ Pauli matrices, for example,
\begin{equation}\label{e02}
\hat S^z =
\begin{pmatrix}\hat \psi^{\dag}_{+1}&\hat \psi^{\dag}_0&\hat \psi^{\dag}_{-1}\\\end{pmatrix}
    \begin{pmatrix} 1 & 0 & 0 \\0 & 0 & 0 \\0 & 0 & -1\end{pmatrix}
    \begin{pmatrix}\hat \psi_{+1}\\\hat \psi_0\\\hat \psi_{-1}\end{pmatrix} .
\end{equation}
Within the mean-field approximation, we treat the spin-dependent
interactions as a molecular field except of a particle with itself,
\begin{equation}\label{e03}
-\frac 12 \hat{\bf S}\cdot \hat{\bf S} \approx - \langle \hat{\bf
S}\rangle \cdot \hat{\bf S}
     +\frac 12\langle \hat{\bf S}\rangle \cdot \langle \hat{\bf S}\rangle
   = - {\overline M}\hat S^z + \frac 12 {\overline M}^2 ,
\end{equation}
where $\overline{M}=\langle \hat S^z\rangle$ is the ferromagnetic
order parameter. Then the effective Hamiltonian for the grand
canonical ensemble reads,
\begin{equation}\label{e04}
\hat H-\hat N\mu= \sum_{{\bf k}\sigma}
   [ \epsilon_{\bf k} - \mu - \sigma (h_m + h_e) ] \hat n_{{\bf k}\sigma}
   +\frac{1}{2}\overline{M}^2I_s N ,
\end{equation}
where $\epsilon_{\bf k}$ is the kinetic energy for free particles,
$h_m=I_s\overline{M}$ is called the molecular field, similar to the
Stoner theory for fermion gases\cite{mohn}; $\mu$ is the chemical
potential; $N$ is the total particle number. The grand thermodynamic
potential can be worked out in a standard way,
\begin{eqnarray}\label{e05}
\Omega & = & - k_BT\ln Tr\exp [ - \frac{\hat H - \hat N\mu }{k_BT}]\nonumber\\
&=&-\frac{(k_BT)^{\frac{5}{2}}Vm^{\frac{3}{2}}}{(2\pi\hbar^2)^{\frac{3}{2}}}\sum_\sigma
    f_{\frac{5}{2}}\left( \frac{\mu+\sigma h}{k_BT} \right)
    +\frac{1}{2}\overline{M}^2I_s N,
\end{eqnarray}
where $h=h_m+h_e$, $m$ is the mass of particle, and $f$ is the
polylogarithm function defined by
\begin{equation}\label{e06}
f_n(x)\equiv\sum_{k=1}^{\infty}\frac{(e^x)^k}{k^n}~,
\end{equation}
where $x\leq 0$. The mean-field self-consistent equations are
derived from the grand thermodynamic potential,
\begin{subequations}\label{e-sf}
\begin{eqnarray}
n &= & - \frac{1}{V} \left( {\frac{{\partial \Omega }}{{\partial \mu
}}} \right)_{T,V}+n_0\nonumber\\
&=&\left(\frac{k_BTm}{2\pi\hbar^2}\right)^{\frac{3}{2}}\sum_\sigma
   f_{\frac{3}{2}}\left( \frac{\mu+\sigma h}{k_BT} \right)+n_0;\\
M &=&  - \frac{1}{V}\left( {\frac{{\partial \Omega }}{{\partial
h_e}}} \right)_{T,V}+n_0\nonumber\\
&=&\left(\frac{k_BTm}{2\pi\hbar^2}\right)^{\frac{3}{2}}\left[
   f_{\frac{3}{2}}\left(  \frac{\mu+ h}{k_BT}\right) \right. \nonumber\\
    &&\left. -f_{\frac{3}{2}}\left(   \frac{\mu-h}{k_BT}\right) \right]+n_0~;
\end{eqnarray}
\end{subequations}
where $n$ is the density of particles, $n_0$ is the density of
condensed one and $M\equiv\frac{N\overline M}{V}$ is the
magnetization. $n_0$ is zero unless the temperature is below the BEC
point $T_c$.

\section{The free energy and specific heat}
In our previous investigations, we showed that the system exhibits
two phase transitions, the Bose-Einstein condensation (BEC) and the
ferromagnetic transition~\cite{gu1}. The condensation temperature
$T_C$ and the Curie temperature $T_F$ are calculated by solving the
self-consistent equations. We find that $T_F$ is never below $T_C$
for all systems with a finite ferromagnetic exchange ($I_s\ne 0$).

However, one can get another solution to the Eqs. (\ref{e-sf}), with
$M=0$ at all temperatures. It means the system does not undergo a
ferromagnetic transition at all, but remains in paramagnetic (PM)
state at low temperatures. Actually, whether there exists a Curie
point in the ferromagnetic Bose gases is still a controversial
question. Some researchers suppose that the Bose gas can not be
magnetized spontaneously at low temperatures even if the
ferromagnetic exchange is present~\cite{isoshima2}.

\begin{figure}
\includegraphics[width=0.35\textwidth,keepaspectratio=true]{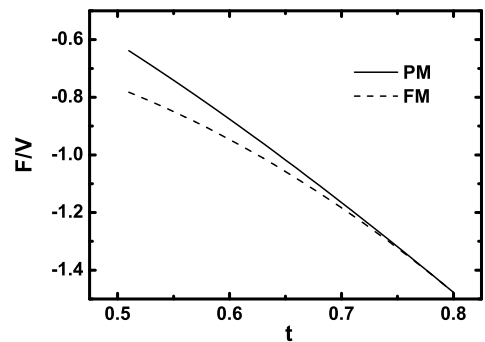}
\caption{Free energies of the FM and PM states with the
ferromagnetic coupling $I=1$. The two curves cross at the
temperature $t_f\approx 0.80$, which is just the FM transition
point.} \label{FE}
\end{figure}%

In order to single out the physically correct solution, one has to
compare the free energy of the the ferromagnetic (FM) state and the
PM state. The relation between the free energy and the grand
thermodynamic potential has the form:
\begin{equation}\label{e08}
  F = \Omega  + N\mu~.
\end{equation}
For computational convenience, the temperature $T$ and exchange
interaction $I_s$ are re-scaled, as did in Ref. [8], by the
following formula:
$t={[3\zeta(\frac{3}{2})]}^{-\frac{2}{3}}{T}/{T_0}$ and
$I={[3\zeta(\frac{3}{2})]}^{-\frac{2}{3}}I_s /( k_B T_0 )$, where
$$T_0 = \frac{1}{{k_B}}\left(\frac{n}{{3\zeta(\frac{3}{2})}}\right)^{\frac{2}{3}}
     \left(\frac{{2\pi \hbar ^2 }}{m}\right)$$
is the condensation temperature of ideal spin-1 Bose gas.
Hereinafter, all the numerical results are obtained by setting
$n=k_B=\frac{2\pi\hbar^2}{m}=1$. Figure 1 shows the free energy of
unit volume for the gas with $I=1.0$.  It shows clearly that the
free energy of FM state is lower than that of the PM state at the
low temperature region, which demonstrates that the FM state should
be more stable than PM state. Therefore, the low temperature state
has a spontaneous magnetization. In experiments, the total spin of
the ferromagnetic spinor condensate is observed to be conserved,
which is called the {\sl spin conservation rule} in some
literatures~\cite{schma,chang}. However, the spin conservation rule
holds only globally, not locally. In the theoretical treatment of
Ref. [20], the spin conservation rule is imposed by introducing a
lagrangian multiplier. It is overconstrained in some sense, so that
the spontaneous magnetization can not be established. Recent
experiments and theories indicate some domain structures should be
formed and each domain is magnetized~\cite{gu2, mur,Berkeley}, where
the conservation law for the total spin can be restored naturally.

The FM transition is induced by the FM coupling and the transition
temperature is about $t_f\approx 0.8$ for the Bose gas with $I=1.0$.
When the temperature goes down further, the BEC then occurs, which
is the intrinsic phase transition of Bose gases. To demonstrate
different features of the two transitions, we now calculate the
specific heat of unit volume,
\begin{equation}\label{e09}
  C =\frac{1}{V} \left( {\frac{{\partial U }}{{\partial T}}} \right)_{B,V}~,
\end{equation}
where $U$ is the internal energy
\begin{eqnarray}\label{e10}
U &=& F-TS = \Omega  - T(\frac{{\partial \Omega }}{{\partial T}})
     + N\mu\ \\\nonumber
  &=& \frac{3V(k_BT)^{\frac{5}{2}}m^{\frac{3}{2}}}{2(2\pi\hbar^2)^{\frac{3}{2}}}\sum_\sigma
     f_{\frac{3}{2}}\left(  \frac{\mu+\sigma h}{k_BT} \right) ~.
\end{eqnarray}
As shown in Fig. 2, for the system with $I=1.0$, the specific heat
exhibits a jump discontinuity at $t_f\approx 0.8$, from the PM state
to the FM state. This is a characteristic feature of the Landau-type
of second-order phase transition. And similar behaviors have been
observed in the specific heat of ferromagnetic insulators or
itinerant-fermion ferromagnets~\cite{mohn,moriya}. The BEC occurs at
$t_c\approx 0.5$, where the specific exhibits a bend. But specific
heat is continuous at the BEC point, similar to that of a free Bose
gas. The results indicate that the critical behaviors are different
at the two transition points on the mean-field level.
\begin{figure}
\includegraphics[width=0.35\textwidth,keepaspectratio=true]{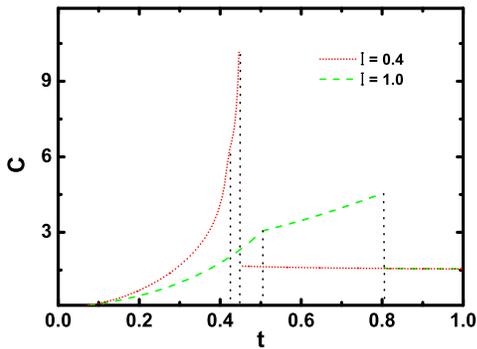}
\caption{Specific heats of spinor Bose gases with the coupling
$I=1.0$ and $0.4$. The dotted vertical lines serve to guide the eye
to see the transition points.} \label{Cv}
\end{figure}

\section{The Curie-Weiss law}

For a ferromagnet, the susceptibility above the Curie point is of
special interest. As already studied, the susceptibility is well
described by Curie-Weiss law both in the insulating ferromagnet and
the itinerant-electron ferromagnet. In this section we calculate the
susceptibility for the itinerant-boson ferromagnet.

The susceptibility can be derived from Eqs. (\ref{e-sf}).
Differentiating both sides of the two equations and removing the
term $d\mu$, the following equation are deduced,
\begin{equation}\label{e11}
dM = fd\left( {\frac{{I_s\overline{M}+h_e}}{k_BT}} \right),
\end{equation}
where
\begin{eqnarray}\label{e12}
f &=& \left(\frac{k_BTm}{2\pi\hbar^2}\right)^{\frac{3}{2}}
     \left[ {f_{\frac{1}{2}} \left( {\frac{{\mu  + h}}{k_BT}} \right)
     + f_{\frac{1}{2}} \left( {\frac{{\mu  - h}}{k_BT}} \right)}
     \right]\nonumber \\
   &&-\left(\frac{k_BTm}{2\pi\hbar^2}\right)^{\frac{3}{2}}
    \frac{{\left[ {f_{\frac{1}{2}} \left( {\frac{{\mu  + h}}{k_BT}}
    \right) - f_{\frac{1}{2}} \left( {\frac{{\mu  - h}}{k_BT}} \right)}
    \right]^2 }} {{\sum\limits_\sigma {f_{\frac{1}{2}}
    \left( {\frac{{\mu  + h\sigma }}{k_BT}} \right)} }}~.
\end{eqnarray}
Above the Curie point, the magnetization $M$ (then
$h=I_s\overline{M}+h_e$) diminishes correspondingly when the
external field $h_e$ tends to zero. So the second term in the above
equation is omitted and then  $f$ has a simple form:
\begin{equation}\label{e13}
f\approx2 \left(\frac{k_BTm}{2\pi\hbar^2}\right) ^{\frac{3}{2}}
    f_{\frac{1}{2}} \left( \frac{\mu }{k_BT} \right).
\end{equation}
Thus the zero-field susceptibility of unit volume is given by
\begin{equation}\label{e14}
\chi  = \left( {\frac{{\partial M}}{{\partial h_e}}} \right)_{T,V} =
\frac1{k_BTf^{-1} - n^{-1}I_s}~.
\end{equation}
The susceptibility $\chi$ is a function of the coupling $I_s$ and
temperature $T$. Figure 3 shows $1/\chi$ and $\chi$ versus $I$ at
different given temperatures. As shown in the inset of Fig. 3, the
susceptibility becomes larger as the coupling $I$ increasing. It is
physically reasonable since the the system with larger $I$ can be
magnetized more easily. At a given temperature, $\chi$ diverges as
$I$ approaches a critical value. It is worth noting that the inverse
of the susceptibility is in a good linear relationship with the
coupling.

\begin{figure}
\includegraphics[width=0.35\textwidth,keepaspectratio=true]{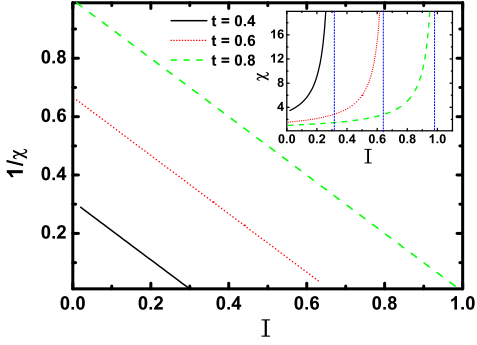}
\caption{Magnetic susceptibilities versus ferromagnetic couplings of
spinor Bose gases at temperature $t=0.4, 0.6$ and $0.8$.}
\label{X-I}
\end{figure}

The susceptibility versus temperature is shown in Fig. 4. One can
immediately find that the susceptibility meet quite well with the
Curie-Weiss law in a very large temperature region. Seeing that the
Curie-Weiss law is very difficult to be derived for the
itinerant-{\sl fermion} ferromagnet, it is really surprising that we
get it for the itinerant-{\sl boson} ferromagnet just based on the
mean-field approximation.

\begin{figure}
\includegraphics[width=0.35\textwidth,keepaspectratio=true]{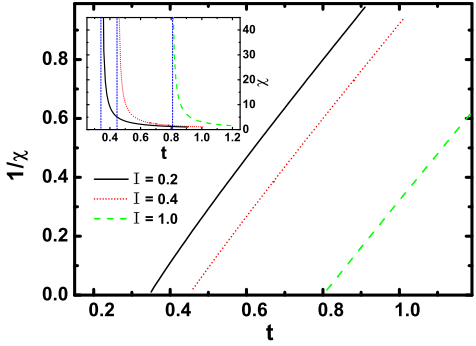}
\caption{Magnetic susceptibilities versus temperatures of spinor
Bose gases on coupling $I=0.2, 0.4$ and $1.0$. } \label{X-T}
\end{figure}

In order to discuss the Curie-Weiss law in a more explicit way, we
proceed to carry out a semi-analytical calculation to deduce the
linear dependence of $1/\chi$ on the temperature. The first step is
to analyze the temperature dependence of $f$. It is quite
complicated, because the chemical potential $\mu$ is an implicit
function of the temperature. We consider a limit case that the
parameter $I_s$ is quite small, when $T_f$ is close to $T_c$. So
$\mu$ is close to zero in the vicinity of $T_f$. According to the
asymptotic behavior of the polylogarithm function: $f_{\frac{3}{2}}
(x) \approx \zeta(\frac{3}{2})-2\sqrt {\pi x}$ and $f_{\frac{1}{2}}
(x) \approx \sqrt {\pi/x}$ as $x\to 0^-$, we get the following
equations from Eqs. (7a) and (13) respectively,
\begin{equation}\label{e15}
n\approx3\left(\frac{k_BTm}{2\pi\hbar^2}\right)^{\frac{3}{2}}\left[f_{\frac{3}{2}}(0)-2\sqrt{-\frac{\pi\mu}{k_BT}}\right].
\end{equation}
and
\begin{equation}\label{e16}
f \approx 2 \left(\frac{k_BTm}{2\pi\hbar^2}\right) ^{\frac{3}{2}}
    \sqrt { -\frac{k_BT\pi}{\mu} }~.
\end{equation}
Substitute Eqs. (\ref{e15}) and (\ref{e16}) into Eq. (\ref{e14}), we
get
\begin{equation}\label{e17}
\chi^{-1} = \frac{nk_B^{-2}}{12\pi}
    \left(\frac{m}{2\pi\hbar^2}\right)^{-3} T^{-\frac{1}{2}}
    \left(T_0^{-\frac{3}{2}}-T^{-\frac{3}{2}}\right) - n^{-1}I_s~.
\end{equation}
In the vicinity of $T_F$ which is only slightly larger than $T_0$,
Eq. (\ref{e17}) could be further simplified to
\begin{eqnarray}\label{e18}
\chi^{-1} &\approx& \frac{nk_B^{-2}}{8\pi} \left(\frac{m}{2\pi\hbar^2}\right)^{-3}T_0^{-3}
     \left(T-T_0\right)-n^{-1}I_s\nonumber \\
   &=&\frac{9\zeta^2(\frac{3}{2})}{8\pi}n^{-1}k_B
   \left[T-\left(T_0 + \frac{8\pi}{9\zeta^2(\frac{3}{2})k_B}I_s\right) \right]~.
\end{eqnarray}
Thus the effective FM transition temperature is defined as
$$T_f=T_0 + \frac{8\pi}{9\zeta^2(\frac{3}{2})k_B}I_s.$$
So far the Curie-Weiss law is derived. We note that the
derivation is only valid in small $I_s$ cases.

In the high temperature limit, one can also easily prove that
$\chi^{-1}$ is linearly dependant on $T$. In this case, $-\frac \mu
{k_BT}$ has a quite large value, so that
$$f_{\frac{1}{2}}\left({\frac \mu {k_BT}}\right) \approx
f_{\frac{3}{2}}\left({\frac \mu {k_BT}}\right) \approx e^{\frac \mu
{k_BT}}$$ according to Eq. (6). Combining Eqs. (7a), (13) and (14),
it yields
\begin{equation}\label{e19}
\chi^{-1}=n^{-1}(k_BT-I_s).
\end{equation}
We estimate this equation holds in the range of $t\gtrsim 10$.

\section{Summary}

In summary, we calculate thermodynamic quantities of the spinor Bose
gas with ferromagnetic interactions. Such kind of investigations has
already been performed intensively for the ferromagnetic fermions,
while few as yet for bosons. Based on a mean-field approximation, we
show that the system undergos a ferromagnetic phase transition
first, then the Bose-Einstein condensation with the temperature
decreasing. The specific heat shows a jump discontinuity at the
Curie point and a bend at the Bose-Einstein condensation
temperature, indicating that critical behaviors are different near
the two transition. The more surprising result is that the
mean-field theory yield the magnetic susceptibility which satisfies
perfectly the Curie-Weiss law over a wide range of temperature.

This work is supported by the National Natural Science Foundation of
China (Grant No. 10504002), the Fok Yin-Tong Education Foundation,
China (Grant No. 101008), and the Ministry of Education of China
(Grant No. NCET-05-0098).


\end{document}